\documentclass[journal=cmatex,manuscript=article,layout=twocolumn]{achemso}
\usepackage{array}
\usepackage{datetime}
\usepackage{textcomp}
\usepackage{gensymb}
\usepackage{graphicx}
\usepackage[hidelinks]{hyperref}
\usepackage{multirow}
\usepackage{natbib}
\usepackage{mciteplus}
\usepackage{soul}
\usepackage[fleqn]{amsmath}
\newcommand\ce[1]{\ensuremath{\mathrm{#1}}}

\usepackage{xcolor} 

\setkeys{acs}{usetitle=true} % titles in references
\SectionNumbersOn

\title{Phase Behavior and Ion Transport in Lithium–Niobium–Tantalum Oxide Alloys}

\author{Hengning Chen}
\affiliation{Department of Materials Science and Engineering, National University of Singapore, 9 Engineering Drive 1, 117575, Singapore}

\author{Zeyu Deng}
\affiliation{Department of Materials Science and Engineering, National University of Singapore, 9 Engineering Drive 1, 117575, Singapore}

\author{Gopalakrishnan Sai Gautam}
\affiliation{Department of Materials Engineering, Indian Institute of Science, 560012 Bangalore, India}

\author{Yan Li}
\affiliation{Department of Materials Science and Engineering, National University of Singapore, 9 Engineering Drive 1, 117575, Singapore}

\author{Pieremanuele Canepa}
\affiliation{Department of Electrical and
Computer Engineering, University of Houston, Houston,
Texas 77204, USA}
\alsoaffiliation{Texas Center of Superconductivity, University of Houston, Houston,
Texas 77204, USA}
\affiliation{Department of Materials Science and Engineering, National University of Singapore, 9 Engineering Drive 1, 117575, Singapore}
\email{pcanepa@uh.edu}
\begin{document}
%%%%%%%%%%%%%%%%%%%%%%%%%%%%%%%%%%%%%%%%%%%%%%%%%%%%%%%%%%%%%%%%%%%%%%%%%%%%

\maketitle

\begin{abstract}

Lithium niobate-tantalate mixtures have garnered considerable interest for their ability to merge the desirable properties of both end members, enabling diverse high-value applications, such as high-performance faradaic capacitors, non-linear optics, and protective coatings in rechargeable batteries. While numerous studies on the application of \ce{LiNb_xTa_{1-x}O_3} exist, the phase behavior and properties of \ce{Li_3Nb_xTa_{1-x}O_4} remain largely unexplored. In this work, we employ a multiscale approach that encompasses first-principles phonon calculations, cluster expansion, and Monte Carlo simulations to derive the temperature-composition phase diagram for \ce{Li_3Nb_xTa_{1-x}O_4}. Our findings reveal the critical role of vibrational entropy in accurately predicting phase stability, which promotes the solubility of Nb in \ce{Li_3TaO_4} while suppressing the miscibility of Ta in \ce{Li_3NbO_4}. Additionally, we demonstrate that Nb/Ta mixing offers a promising avenue for tailoring the Li-ion conductivities of \ce{Li_3Nb_xTa_{1-x}O_4}. On the technical side, we demonstrated the importance of including vibrational entropy effects explicitly in Monte Carlo simulations dealing with multicomponent systems, beyond simple binary mixtures.  On the application side, this study provides fundamental insights into the phase behavior and Li-ion transport properties of  \ce{Li_3Nb_xTa_{1-x}O_4}, paving the way for its potential applications in energy storage and other fields. 
\end{abstract}

%%%%%%%%%%%%%%%%%%%%%%%%%%%%%%%%%%%%%%%%%%%%%%%%%%%%%%%%%%%%%%%%%%%%%%%%%%%%

\section{Introduction}
Functional oxides are employed in a myriad of applications and technologically relevant devices.\cite{FernandezGarcia2004, Dai2003Novel, Yuan2014Mixed} These materials are often used as protective coating materials in energy storage and conversion applications.\cite{Culver2019CoatingSSB, Tan2021NatureNano, li2006cathode, guan2020recent} In particular, the increasing demand for high-performance and safe energy storage devices has driven significant research into advanced lithium(Li)-ion batteries (LIBs) and all-solid-state batteries (ASSBs) with large gravimetric (volumetric) energy and power densities.\cite{Janek2016SSB,Goodenough2009challenges} The improvement of the safety of these battery devices is not secondary. Unfortunately, LIB devices use flammable electrolytes and are plagued by a persistent cathode-electrolyte interface degradation, hindering these cells' overall efficiencies and lifespans.\cite{Liu2015Ni-rich,wang2015surface} Li-metal anode-based ASSBs offer safer and potentially higher energy and power densities compared to commercial LIBs, but face issues with electrode-electrolyte compatibility, inadequate physical contact between particles of solid electrolytes and the active materials, as well as metal ingress in the solid electrolyte separator, and often termed as dendrite nucleation and growth.\cite{Janek2016SSB, Xie2024} Coating technologies have emerged as one of the most crucial strategies to mitigate some of these challenges in LIBs and ASSBs.\cite{li2006cathode,guan2020recent, Meyerson2020,kim2019stabilizing} Functioning as physical and chemical protective layers, coating materials can mitigate interface instabilities, modulate resistance to ion and electron transport, and improve the overall cell performance.\cite{Famprikis2019,Kato2016LNcoating, Culver2019CoatingSSB}
   
Prior experimental works have demonstrated the great potential of both amorphous and crystalline Li-Nb-O and Li-Ta-O coatings for positive electrode materials in LIBs and ASSBs.\cite{Ohta2007amorphousLN,Xin2021LNL3Ncoating,Li2023L3NcoatingPANI,You2024LiTaO3coating} Amorphous \ce{LiNbO_3} and \ce{LiTaO_3} materials exhibit adequate ionic conductivities ($\sim$\ce{10^{-6}-10^{-7}}~S~\ce{ cm^{-1}}) at room temperature, which might be attributed to their high porosity (approximately 22 $\%$ free volume).\cite{Glass1978Amorphous,Uhlendorf2017amorphousLNLT,Hueger2023LNpolymorphic} Crystalline niobate and tantalate coatings reveal a multiphasic nature comprising \ce{LiMO_3} and \ce{Li_3MO_4} (M=Nb, Ta) phases,\cite{Xin2021LNL3Ncoating,Chen2023NTcoating,Nyman2008, GuozhongLu2021} with {\ce{Li_3MO_4}} actively involved in Li-ion transport due to its defect-rich landscape.\cite{Kim2023L3Tdefect,Liao2013L3NGoodenough,Chen2023NTcoating}

In 2015, utilizing physical vapor deposition approach, \citeauthor{Yada2015} conducted a high-throughput exploration of the phase-space of the \ce{LiNb_xTa_{1-x}O_3} system.\cite{Yada2015} They reported enhanced Li-ion conductivity in partially crystallized \ce{Li_{56}Nb_{22}Ta_{22}} (4.2$\times$\ce{10^{-6}} S \ce{cm^{-1}}) and higher permittivity of 165 when measured at 254 kHz compared to pure \ce{LiNbO_3} (1.8$\times$\ce{10^{-6}} S \ce{cm^{-1}} and 95, respectively). Generally, \ce{LiNb_xTa_{1-x}O_3} demonstrates improved Li-ion diffusion and reduced charge transfer resistance at interfaces within ASSBs.\cite{Zhangwenbo2017,LiuAl2023} As the concentration of Nb ($x$) increases, the ionic conductivities of \ce{LiNb_xTa_{1-x}O_3} first increase and then decline, culminating in peak conductivity when $x$=0.5. This finding is further supported by Wang et al.\cite{Wang2020Unveiling}, wherein the ionic conductivity of partially crystallized \ce{LiNb_{0.5}Ta_{0.5}O_3} reaches 38.7$\times$\ce{10^{-6}} S \ce{cm^{-1}}.

Due to the isovalent electronic configuration and nearly identical ionic radii (64 pm) of \ce{Nb^{5+}} and \ce{Ta^{5+}} when six-coordinated,\cite{Shannon1976} tantalum can easily substitute on Nb sites without introducing intrinsic defects, such as vacancies or interstitials. Mixing \ce{LiNbO_3} with \ce{LiTaO_3} phases exhibits complete solid solubility across the entire composition range.\cite{Bartasyte2012, Shimura1977}

Despite the extensive research on the \ce{LiNb_xTa_{1-x}O_3} system,\cite{Huband2017,Suhak2021,Roshchupkin2020,Ruesing2016vibrationalLNT} understanding the phase behavior of \ce{Li_3Nb_xTa_{1-x}O_4} is crucial to the design of superior crystalline coating materials, which remains unexplored. 
To bridge this knowledge gap, this study examines the phase stability of \ce{Li_3Nb_xTa_{1-x}O_4} by providing a complete thermodynamic picture of this system that incorporates both configurational and vibrational entropy contributions. %We recognize the significant impact of lattice vibrations on thermodynamic stability.\cite{Walle2002vibration}

Employing a multiscale approach parametrized on accurate first-principles calculations, cluster expansion methods, phonon calculations, and atomistic Monte Carlo (MC) simulations, we investigate the mixing of Nb and Ta in the pseudo-binary \ce{Li_3Nb_xTa_{1-x}O_4} system. Our findings reveal the significant contribution of lattice vibrations to the phase transition in \ce{Li_3Nb_xTa_{1-x}O_4}, resolving discrepancies between theoretical predictions considering only configurational entropy and experimental observations. This discrepancy can be attributed to the distinct vibrational characteristics of Nb-O and Ta-O bonds and differences in atomic arrangements inherent to the \ce{Li_3NbO_4} and \ce{Li_3TaO_4} local structures. We observe that the Li-ion conductivity of \ce{Li_3Nb_xTa_{1-x}O_4} increases with the degree of cation mixing within both $\alpha$- and $\beta$-phases. This work provides a comprehensive understanding of the phase behavior and conduction properties of \ce{Li_3Nb_xTa_{1-x}O_4}, paving the way for the design of advanced coating materials for energy storage and conversion applications.

\section{Methods}

\subsection{First-Principles Calculations}
\label{subsec:abinitio}

We conducted density functional theory (DFT) calculations as implemented in the Vienna \emph{ab initio} Simulation Package.\cite{Kresse1996VASP} In DFT, the unknown exchange-correlation energy was approximated by the meta-GGA functional \ce{r^2SCAN}.\cite{Furness2020r2scan} To systematically explore a diverse range of Nb/Ta mixing configurations in \ce{Li_3Nb_xTa_{1-x}O_4}, we generated all possible orderings in supercell models, accommodating up to 16 formula units (128 atoms) for the cubic phase of \ce{Li_3NbO_4} and the monoclinic phase of \ce{Li_3TaO_4}. At each composition in the \ce{Li_3Nb_xTa_{1-x}O_4} tie-line, a maximum of 1,000 distinct Nb-Ta orderings with the lowest Ewald energy were initially stored,\cite{Ewald1921, Ong2013pymatgen} followed by the selection of only symmetrically inequivalent structures. The computational details of \ce{Li_3Nb_xTa_{1-x}O_4} orderings relaxation follow the prescriptions used in our previous study,\cite{Chen2023NTcoating} with DFT total energy integrated over equal $k$-point densities consistent across different supercell models. 

\subsection{Cluster Expansion} \label{CE}
We applied the cluster-expansion (CE) formalism to the Nb and Ta distributions in the \ce{Li_3Nb_xTa_{1-x}O_4} system. Based on the approach introduced by \citeauthor{Sanchez1984cluster},\cite{Sanchez1984cluster} the CE model connects the effective cluster interactions, $V_{\alpha}$, in Eq.~\ref{eq:ce} with the configurational mixing energies ($\mathrm{E_{mix}(\ce{conf})}$ of Eq.~\ref{Emix}) of distinct Nb/Ta orderings calculated by DFT on supercell models. The fitting of the CE was performed using the cluster-assisted statistical mechanics (CASM) package,\cite{CASMDevelopers2017,Puchala2013ZrOpd,Puchala2023CASM} where $\mathrm{E_{mix}}(\text{conf})$ depends on an expansion of different cluster functions (Eq.~\ref{eq:ce}):
{\footnotesize
\begin{equation}
\label{eq:ce}
   E_{mix}(\text{conf}) =\underset{\sigma}{\Sigma} \underset{\alpha}{\Sigma}V_{\alpha}\overline{\Pi_{\alpha}}(\vec{\sigma}); \;\;  \vec{\sigma} = (\sigma_1, \sigma_2, \cdots, \sigma_i)
\end{equation}
}
where $\vec{\sigma}$ are configuration vectors, consisting of occupation variables, $\sigma_i$ = --1 or 1, representing the Nb or Ta occupancies of specific crystallographic sites, $i$. 
$V_\alpha$ refers to the effective cluster interactions (ECIs), including the multiplicity of symmetrically equivalent clusters, $\alpha$. In the \ce{Li_3Nb_xTa_{1-x}O_4} systems, pairs, triplets, and quadruplets cluster interactions are truncated at maximum radii of 12\AA, 7\AA, and 5\AA, respectively. To increase the sparsity of the ECI solutions, we applied the compressive sensing algorithm to determine the optimal set of ECIs. In particular, we employed the hyperparameter $\alpha = 1 \times 10^{-6}$ to solve the least absolute shrinkage and selection operator problem.\cite{Nelson2013CEalgorithum}

\subsection{Vibrational Entropy Contributions in \ce{Li_3Nb_xTa_{1-x}O_4}}

From two \ce{Li_3Nb_xTa_{1-x}O_4} substitution systems, i.e.\ the $\alpha$-phase and the $\beta$-phase, 10 ground-state structures at selected compositions $x$\,=\,0, 0.25, 0.5, 0.75, and 1 were chosen for harmonic phonon calculations obtained with Phonopy and VASP.\cite{Togo2015phonopy} In these calculations (on 32-atom large cells), stricter convergence criteria were imposed to accept structure convergence: $10^{-8}$ eV for the DFT total energy and $10^{-3}$ eV/{\AA} for the Hellmann-Feynman forces. Then, 2$\times$2$\times$2 supercells (256 atoms) were generated for finite difference calculations. Paths in the phonon band structure plots were obtained with SeeK-path by \citeauthor{Hinuma2016seekpath}\cite{Hinuma2016seekpath} Phonon band structures and phonon density of states are plotted in the SI (Figure S13-14). From these calculations, we can extract the zero-point energies and the integrated values of vibrational free energies, which are then incorporated in the Monte Carlo simulations.

\subsection{Monte Carlo Simulations}\label{semi-gcMC}
To include effects of vibrational free energy, estimated values of $\mathrm{F_{mix}(vib,x,T)}$ at specific compositions and temperatures are passed to the Monte Carlo routines in a modified version of the CASM code,\cite{GitHubCASM} where we incorporated directly $\mathrm{dF_{mix}(vib)}$ into the Metropolis step. 

For substitution in both $\alpha$- and $\beta$-phases, we carried out semi-grand canonical MC simulations based on 10$\times$10$\times$10 model supercells (with 4,000 f.u.) by performing separate chemical potential ($\mu$) and temperature (T) MC scans.\cite{Walle2002MCintegration} First, the scan started from T=10 K and up to T=1,000 K with a step of $\Delta$T=10 K in the range of $\mu$=$\left[-1.0, 1.0\right]$ eV/f.u. ($\Delta \mu$=0.01). Then at every temperature value, $\mu$ was scanned in both the forward (\ce{Li_3TaO_4}, $x$=0 $\rightarrow$ \ce{Li_3NbO_4}, $x$=1, i.e. from $\mu$ = --1.0 to +1.0 eV/f.u.) and the backward ($x$=1 $\rightarrow$ $x$=0, from $\mu$ = +1.0 to --1.0 eV/f.u.) directions with a step size $\delta\mu$=$\pm$0.01 eV/f.u. With these settings, we explored two distinct phase diagrams, one for the $\alpha$ phase and the other for the $\beta$ phase. The complete phase diagram was obtained by comparing the semi-grand canonical potentials of the two phases across the entire composition range to identify the stable structure. Details of the integration methods are described in the SI.

\subsection{Li$^+$ Mobility with Machine Learning Molecular Dynamics}\label{MTP-MD}

We started from \ce{Li_3Nb_xTa_{1-x}O_4} configurations (2$\times$2$\times$2 supercells, 32 f.u.) obtained with semi-canonical Monte Carlo simulations of representative compositions at 300 K. After DFT geometry optimizations of these structures (see details in Sec.~{\ref{subsec:abinitio}}), we performed AIMD simulations whithin the low-limit vacancy regime (one Li vacancy, {\ce{Vac_{Li}}}, in Li$_{95}$Ta$_{32(1-x)}$Nb$_{32x}$O$_{128}$) to generate the training sets for fitting the machine learning potential. The plane-wave energy cutoff was set to 400 eV, and employed the Nos\'e-Hoover thermostat. The Newton equation of motion was integrated with a timestep of 0.5 fs. AIMD simulations were executed for all configurations at 600 K, 800 K, 1,000 K, and 1,200 K for 10 ps, incorporating an initial temperature ramping of 0.5 ps. 

The moment tensor potentials (MTP) are individually trained for each Ta/Nb concentration using 32,000 AIMD snapshots, specifically the final 4 ps trajectory of each temperature. During the MTP training, we set the hyperparameter lev$_{\mathrm{max}}$ of 12 and $\mathrm{R_{cut}}$ to be 7 {\AA} to ensure robust fitting and validation accuracies in energies, forces, and stresses. Extended MD trajectories (10 ns) were obtained after a temperature ramping of 10 ps (to reach the target temperature) and a following period of equilibration of 1 ns.

\section{Results}

\subsection{Phase Stabilities of \ce{Li_3Nb_xTa_{1-x}O_4}}

\ce{Li_3NbO_4} and \ce{Li_3TaO_4} crystallize in distinct structure archetypes. The cubic \ce{Li_3NbO_4} (space group:~$I\overline{4}3m$) features Nb sites (8c) forming edge-sharing Nb octahedra arranged in \ce{Nb_4O_{16}} clusters (Figure S4a). The monoclinic \ce{Li_3TaO_4} (space group:~$C2/c$) shows distinct edge-sharing Ta octahedra giving rise to zig-zag chains along the $c$ direction (Figure S4b). Therefore, mixing \ce{Li_3Nb_xTa_{1-x}O_4} requires consideration of both structural archetypes. For simplicity, we denote the substitution of Ta with Nb in monoclinic \ce{Li_3TaO_4} as $\alpha$-phase substitution and the substitution of Nb with Ta in the cubic \ce{Li_3NbO_4} structure as $\beta$-phase substitution in the following context. 

\begin{figure*}[ht!]
    \centering
    \includegraphics[width=\textwidth]{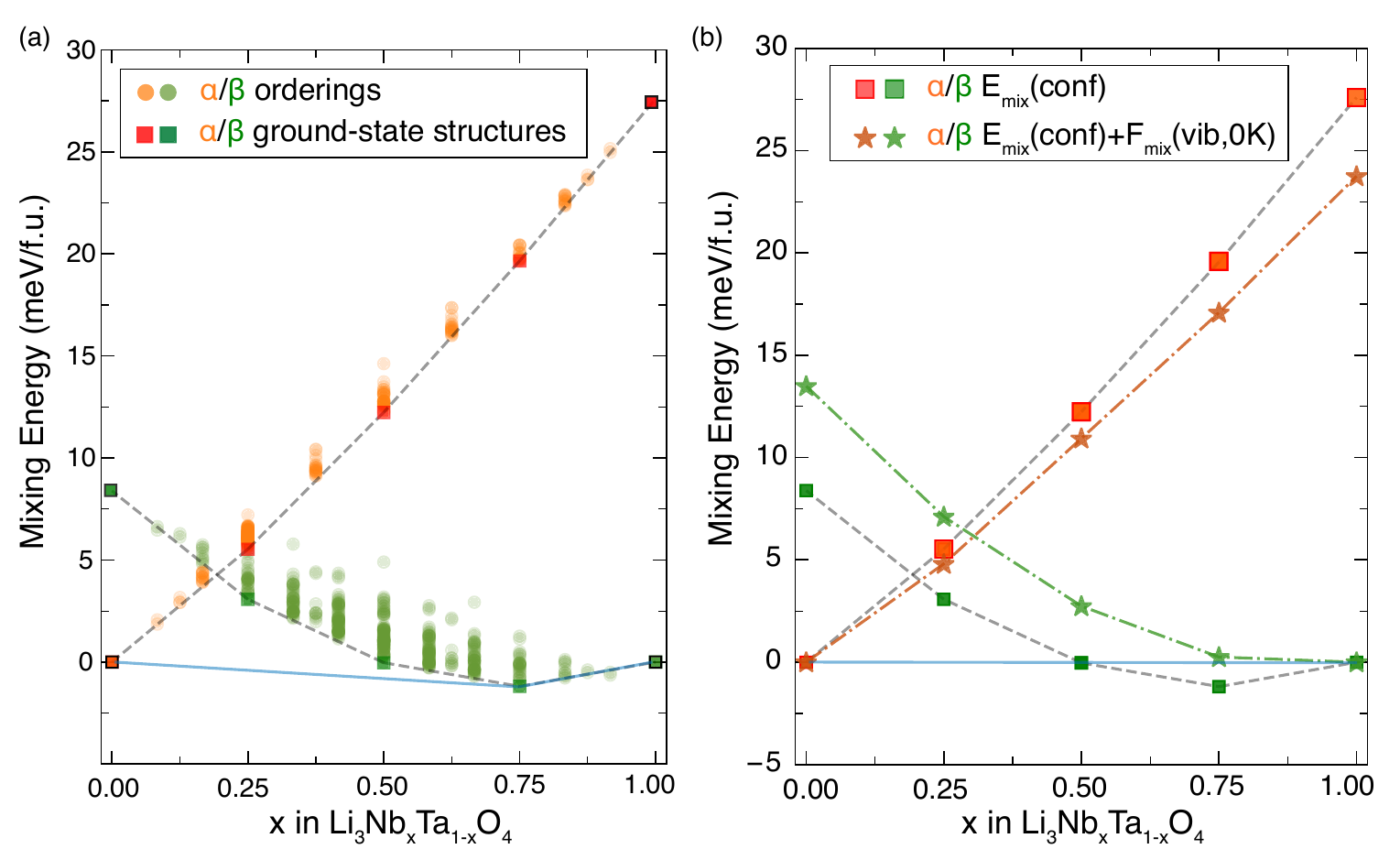}
    \caption{\label{combined_convex_hull}
    Combined mixing energies of $\alpha$-phase and $\beta$-phase \ce{Li_3Nb_xTa_{1-x}O_4} systems: (a) Configurational mixing energies of all Nb/Ta ordering models explored, (b) Configurational and vibrational mixing energies (at 0 K) of only ground-state structures based on DFT calculations. The $\alpha$-phase and $\beta$-phase substituted configurations are displayed with orange and green filled points. Blue solid lines in (a) and (b) represent the global convex hull, while dashed lines show the convex hull envelopes constructed by the ground-state structures of $\alpha$ and $\beta$ phases, separately. In the $\alpha$-phase, the ground-state structures (red squares) are \ce{Li_3TaO_4}($C2/c$), \ce{Li_3Ta_{0.75}Nb_{0.25}O_4}($P1$), \ce{Li_3Ta_{0.5}Nb_{0.5}O_4} ($Cc$), \ce{Li_3Ta_{0.25}Nb_{0.75}O_4}($P1$), \ce{Li_3NbO_4}($C2/c$). In $\beta$-phase, the ground-state structures (green squares) are \ce{Li_3TaO_4}($I\overline{4}3m$), \ce{Li_3Ta_{0.75}Nb_{0.25}O_4}($R3m$), \ce{Li_3Ta_{0.5}Nb_{0.5}O_4} ($Fmm2$), \ce{Li_3Ta_{0.25}Nb_{0.75}O_4}($R3m$), \ce{Li_3NbO_4}($I\overline{4}3m$).
}
\end{figure*}

The mixing energies of \ce{Li_3Nb_xTa_{1-x}O_4} $\alpha$- and $\beta$-phase orderings were calculated as a function of composition ($x$) (Figure~\ref{combined_convex_hull}a). In the $\beta$-phase substitution, the resulting $\beta$-\ce{Li_3TaO_4} (green square at $x$=0) maintains the cubic $I\overline{4}3m$ structure of the $\beta$-\ce{Li_3NbO_4} (green square at $x$=1). All $\beta$-phase configurations feature \ce{(Nb/Ta)_4O_{16}} clusters with a close resamblance to the $\beta$-\ce{Li_3NbO_4} structure. Similarly, in the $\alpha$-phase substitution, both $\alpha$-\ce{Li_3TaO_4} (red square at $x$=0) and $\alpha$-\ce{Li_3NbO_4} (red square at $x$=1) retain the monoclinic $C2/c$ structure. All $\alpha$-phase mixed configurations exhibit zig-zag chains of Nb/Ta octahedra as in \ce{Li_3TaO_4}, though mostly with reduced symmetry. 

From the convex hull in Figure~{\ref{combined_convex_hull}}a, it is clear that the polymorphs generated from substitutions, i.e., $\alpha$-\ce{Li_3NbO_4} and $\beta$-\ce{Li_3TaO_4} are metastable. These results confirm the greater stability of $\beta$-\ce{Li_3NbO_4} and $\alpha$-\ce{Li_3TaO_4}, consistent with experimental observations.\cite{kim2022synthesis,hsiao2010preparation}

To construct the phase diagram for \ce{Li_3Nb_xTa_{1-x}O_4}, we combined the results from both substitution systems, using the stable phases as reference endpoints, $\alpha$-\ce{Li_3TaO_4} and $\beta$-\ce{Li_3NbO_4} in the pseudo-binary tie line, as depicted in Figure \ref{combined_convex_hull}. Thus, the configurational mixing energy $\mathrm{E_{mix}(conf)}$ is calculated as in Eq.~\ref{Emix}: 
\begin{equation}
    \small
    \begin{aligned}
    E_{mix}(\mathrm{conf}) =\ & E(\mathrm{Li_3Nb_{x}Ta_{1-x}O_4}) + \\
    & - (1{-}x)\,E(\alpha\text{-}\mathrm{Li_3TaO_4}) + \\
    & - x\,E(\beta\text{-}\mathrm{Li_3NbO_4})
    \end{aligned}\label{Emix}
\end{equation}
where $E$(\ce{Li_3Nb_xTa_{1-x}O_4}), $E(\alpha$-Li$_3$TaO$_4)$, and $E(\beta$-Li$_3$NbO$_4)$ represent the DFT total energies of each Nb/Ta ordering and the two stable end-member structures at 0 K. As shown in Figure \ref{combined_convex_hull}a, the global convex hull comprises $\alpha$-\ce{Li_3TaO_4}, $\beta$-\ce{Li_3Ta_{0.25}Nb_{0.75}O_4}, and $\beta$-\ce{Li_3NbO_4}. The slightly negative mixing energy (--1.2 meV/f.u.) of \ce{Li_3Ta_{0.25}Nb_{0.75}O_4} suggests a weak thermodynamic driving force for Nb/Ta mixing.  
Other Nb/Ta mixing configurations (green and orange points in Figure~\ref{combined_convex_hull}) are metastable and prone to proportional decompositions --that is phase separation-- into nearby stable phases. 

To investigate the temperature and composition-dependent thermodynamic properties of \ce{Li_3Nb_xTa_{1-x}O_4}, we employed  DFT calculations of many Nb/Ta orderings (see Section~\ref{CE}) to parameterize a cluster expansion (CE) Hamiltonian. Subsequently, the CE Hamiltonian was used to perform semi-grand canonical Monte Carlo simulations probing the effect of temperature on the mixing properties of the \ce{Li_3Nb_xTa_{1-x}O_4} system.

Of the DFT-calculated configurations, 223 and 489 for $\alpha$-phase and $\beta$-phase orderings, respectively, are selected for fitting the CE model (Figure S4a-b). The effective cluster interactions (ECIs) were determined by minimizing the root mean square error (RMSE) between the CE model predictions and all DFT-calculated energies. The leave-one-out cross-validation (LOOCV) score was used to evaluate the predictive accuracy of the CE. The final ECIs for the $\alpha$- and $\beta$-phases in Figures S5 and S6 are obtained with small values of RMSEs $\sim$0.30, 0.33 meV/f.u., and LOOCV scores  $\sim$0.36, 0.50 meV/f.u., respectively. As illustrated in Figures S5 and S6, the cluster expansion analysis reveals that the nearest-neighbor pair interactions between Nb and Ta substitutions dominate the mixing energy landscape, with higher-order triplet and quadruplet clusters contributing negligibly.  

Given the proximity in configurational mixing energies (within 30 meV/f.u.) among the phases defining the 0 K convex hull, the calculated mixing vibrational free energy contributions (--20$\sim$25 meV/f.u.\ in the 0-1000 K range) are likely to influence the relative stability of ground-state structures. As shown in Figure S10, the vibrational free energies exhibit a near-linear relationship as a function of concentration $x$, which follows Vegard's law (except for some minor deviations). A smooth interpolation of the ground state values at each temperature (Figures S11),  allows us to efficiently determine vibrational free energies of \ce{Li_3Nb_xTa_{1-x}O_4} at other concentrations, thus bypassing computationally demanding calculations for all configurations \cite{Walle2002vibration} or more complex uses of the CE paradigm.\cite{Wolverton2000, Shchyglo2008, Zhuravlev2017, Zhuravlev2014}

The mixing vibrational free energy,\cite{Gan2006} $\mathrm{F_{mix}(vib)}$, captures the compositional ($x$) and temperature ($T$) dependence relative to the vibrational free energies of the two stable end members, $\alpha$-\ce{Li_3TaO_4} and $\beta$-\ce{Li_3NbO_4}.
{\footnotesize
\begin{equation}\label{Fmixvib} 
    \begin{aligned}
    F_{mix}(\text{vib,T}) = &F_{vib}(\ce{Li_3Nb_xTa_{1-x}O_4,T}) \\
    & - xF_{vib}(\ce{\beta{-}Li_3NbO_4,T}) \\
    & -(1-x)F_{vib}(\ce{\alpha{-}Li_3TaO_4,T})
    \end{aligned}
\end{equation}
}
where $\mathrm{F_{vib}(Li_3Nb_xTa_{1-x}O_4,T)}$, $\mathrm{F_{vib}(\beta{-}Li_3NbO_4,T)}$, and  $\mathrm{F_{vib}(\alpha{-}Li_3TaO_4,T)}$ are the vibrational free energies of \ce{Li_3Nb_xTa_{1-x}O_4}, \ce{\beta}-phase \ce{Li_3NbO_4}, and \ce{\alpha}-phase \ce{Li_3TaO_4} at specific temperature, T, respectively. At each temperature, we used a cubic-type polynomial form to interpolate values of $F_{mix}(\textrm{vib})$ at intermediate compositions.

\begin{figure*}[ht!]
    \centering
    \includegraphics[width=\textwidth]{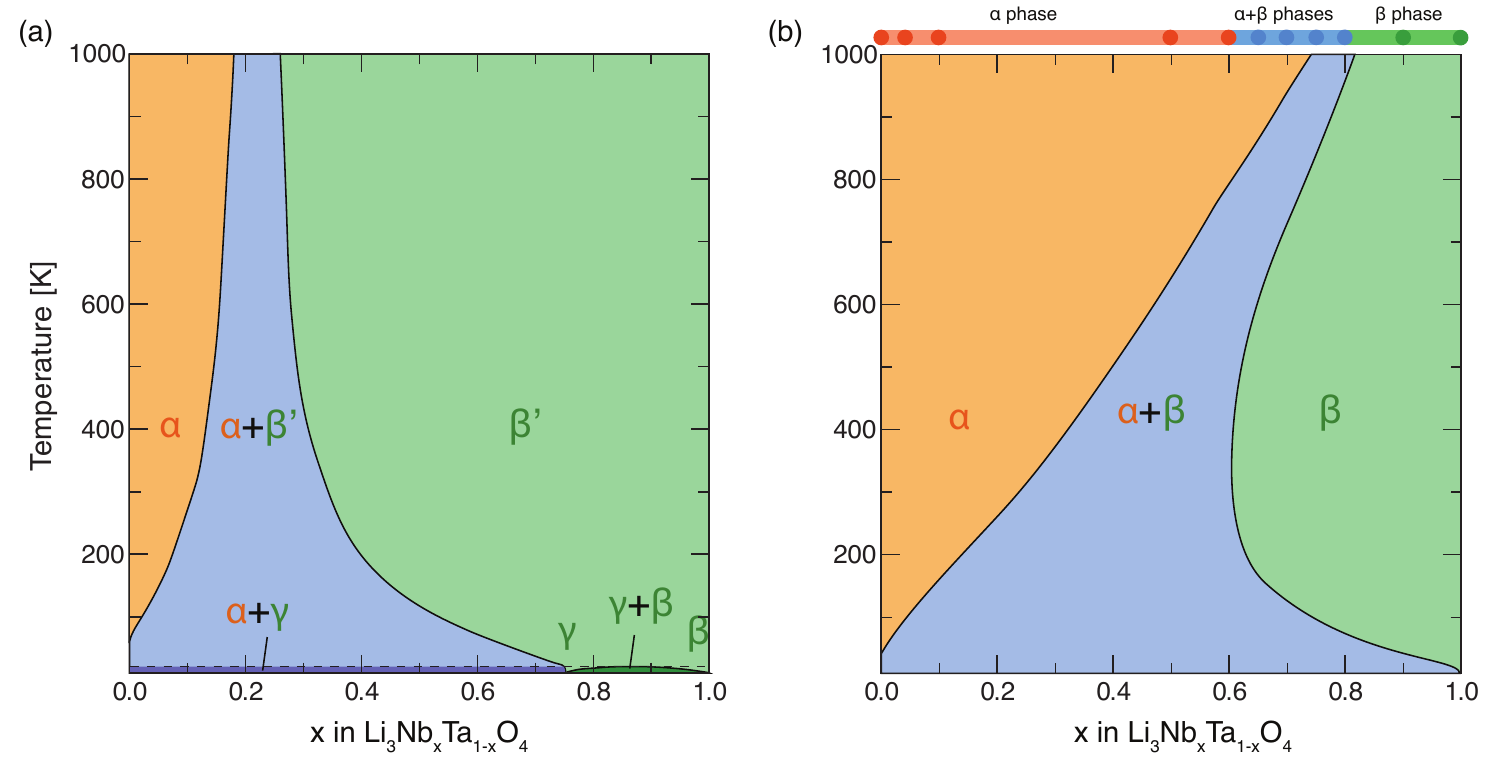}
    \caption{\label{pd_L3NTO4} 
    Computed phase diagrams of \ce{Li_3Nb_xTa_{1-x}O_4} from semi-grand canonical MC simulations at different temperatures (vertical axis) vs.\ variable compositions (horizontal axis), including (a) configurational entropy, (b) configurational and vibrational entropy effects. Colored dots on the top x-axis of (b) show the sequence of stable \ce{Li_3Nb_xTa_{1-x}O_4} phases experimentally synthesized at 1273 K.\cite{Miao1999lumiL3TNO4}}
\end{figure*}

Given the comparable magnitudes of $\mathrm{F_{mix}(vib)}$ and $\mathrm{E_{mix}(conf)}$, the vibrational contributions were incorporated explicitly into the phase diagram construction. The values of $\mathrm{F_{mix}(vib)}$ were then added to $\mathrm{E_{mix}(conf)}$ in Figure \ref{combined_convex_hull}b (at 0 K) and incorporated with the subsequent MC simulations at elevated temperatures (Figure \ref{pd_L3NTO4}b accompained by a detailed description in SI). As shown in Figure S11, $\mathrm{F_{mix}(vib)}$ exhibits negative values in $\alpha$-phase substitution but positive values in $\beta$-phase substitution. This result suggests that the thermal vibrations stabilize the $\alpha$-phase ($C2/c$ \ce{Li_3TaO_4}) substitution while destabilizing the $\beta$-phase ($I\overline{4}3m$ \ce{Li_3NbO_4}) substitution. The total mixing energy, i.e., configurational energy, $\mathrm{E_{mix}(conf)}$ and vibrational free energy, $\mathrm{F_{mix}(vib)}$, at \ce{Li_3Ta_{0.25}Nb_{0.75}O_4} consequently increases to 0.25 meV/f.u., leading to the disappearance of this global minima ($x$=0.75). 

Based on the DFT-calculated orderings and CE-fitted Hamiltonian, the temperature vs.\, Nb composition (x in \ce{Li_3Nb_xTa_{1-x}O_4}) phase diagram was constructed by minimizing the free energies of the $\alpha$-phase and the $\beta$-phase substitutions from semi-grand canonical MC simulations. Figure~\ref{pd_L3NTO4}a depicts the phase diagram of \ce{Li_3Nb_xTa_{1-x}O_4} by only considering the configurational entropy of both systems (Eq.~\ref{phi_config}). In contrast, Figure~\ref{pd_L3NTO4}b includes both configurational and vibrational entropies into the MC simulations (Eq.~\ref{phi_conf_vib}). Section~\ref{semi-gcMC} and SI describe in detail the computational approach.

{\scriptsize
\begin{align}
\Phi_{\mathrm{MC}}(\mathrm{conf}) &= E_{mix}(\mathrm{conf}) - TS_{mix}(\mathrm{conf}) - \mu x, \label{phi_config} \\
 \Phi_{\mathrm{MC}}(\mathrm{vib}) &= \Phi_{\mathrm{MC}}(\mathrm{conf}) + \underbrace{E_{mix}(\mathrm{vib}) - TS(\mathrm{vib})}_{F_{mix}({\mathrm{vib}})} \label{phi_conf_vib}; 
\end{align}
}

As shown in Figure \ref{pd_L3NTO4}a, there are three monophasic regions. $\alpha$ phase include configurations generated through the $\alpha$-phase substitution, while $\gamma$, $\beta$, $\beta^{'}$ phases are based on $\beta$-phase substitution. All these single phases are separated by various biphasic regions. $\gamma$ phase, which maintains the $R3m$ space group of \ce{Li_3Ta_{0.25}Nb_{0.75}O_4} and is characterized by the \ce{Nb_3TaO_{16}} clusters, can only exist at very low temperatures ($<$\, 30~K). The biphasic $\gamma+\beta$ (dark green) domain ceases above $\sim$20 K, forming a wide single-phase region ($0.26 <$x$< 1$ at 1000 K) of \ce{Li_3Nb_{x}Ta_{1-x}O_4}, $\beta^{'}$. The dashed black lines differentiate $\alpha+\gamma$ (purple) and $\alpha+\beta^{'}$ (blue) biphasic regions, while the solid black lines denote the phase boundaries. The $\alpha$ region slightly expands ($0 < x < 0.18$) as temperature increases, leaving a narrow two-phase region of the $\alpha+\beta^{'}$ region ($0.18 < x < 0.26$ at 1000 K).

After introducing the vibrational contributions into the phase diagram construction, the phase diagram in Figure~\ref{pd_L3NTO4}b comprises only 3 main phase regions, including the monophasic $\alpha$ and $\beta$ regions and the $\alpha+\beta$ biphasic region. Moreover, as temperature increases, the $\alpha+\beta$ phase region shifts to higher values of niobium content ($0.74 < x < 0.82$), leaving a wide region of the $\alpha$ phase ($0 < x < 0.74$ at 1000 K) but a narrower region of the $\beta$ phase ($0.82 < x < 1$ at 1000 K). This phase transition in Figure~{\ref{pd_L3NTO4}b}, qualitatively agrees with the experimental findings (colored points in Figure \ref{pd_L3NTO4}b). 

\citeauthor{Miao1999lumiL3TNO4} conducted solid-state synthesis of \ce{Li_3Nb_xTa_{1-x}O_4} across the composition region at 1273 K.\cite{Miao1999lumiL3TNO4} Their X-ray diffraction results indicated that \ce{Li_3Nb_xTa_{1-x}O_4} maintains an $\alpha$-phase structure at $0 < x < 0.6$, $\alpha+\beta$ phases at $0.6 < x < 0.8$, and $\beta$-phase at $0.8 < x < 1.0$, in a good agreement with our computational results. A comparable phase transition trend was also reported by \citeauthor{Blasse1965L3NTO4},\cite{Blasse1965L3NTO4} revealing a greater dissolution ratio of Nb in $\alpha$-\ce{Li_3TaO_4} than that of Ta in $\beta$-\ce{Li_3NbO_4} at synthesis temperature of 1173 K. %But as the authors mentioned, their determination of phase distribution --$\alpha$-phase region, $0 < x < 0.2$, dual-phase $\alpha+\beta$ region $0.2 < x < 0.9$, $\beta$-phase region, $0.9 < x < 1.0$ may not be very accurate given the similar ionic radius of Nb$^{5+}$ and Ta$^{5+}$ and volumes of $\alpha$ and $\beta$ phases.

\begin{figure*}[ht!]
    \centering
    \includegraphics[width=\textwidth]{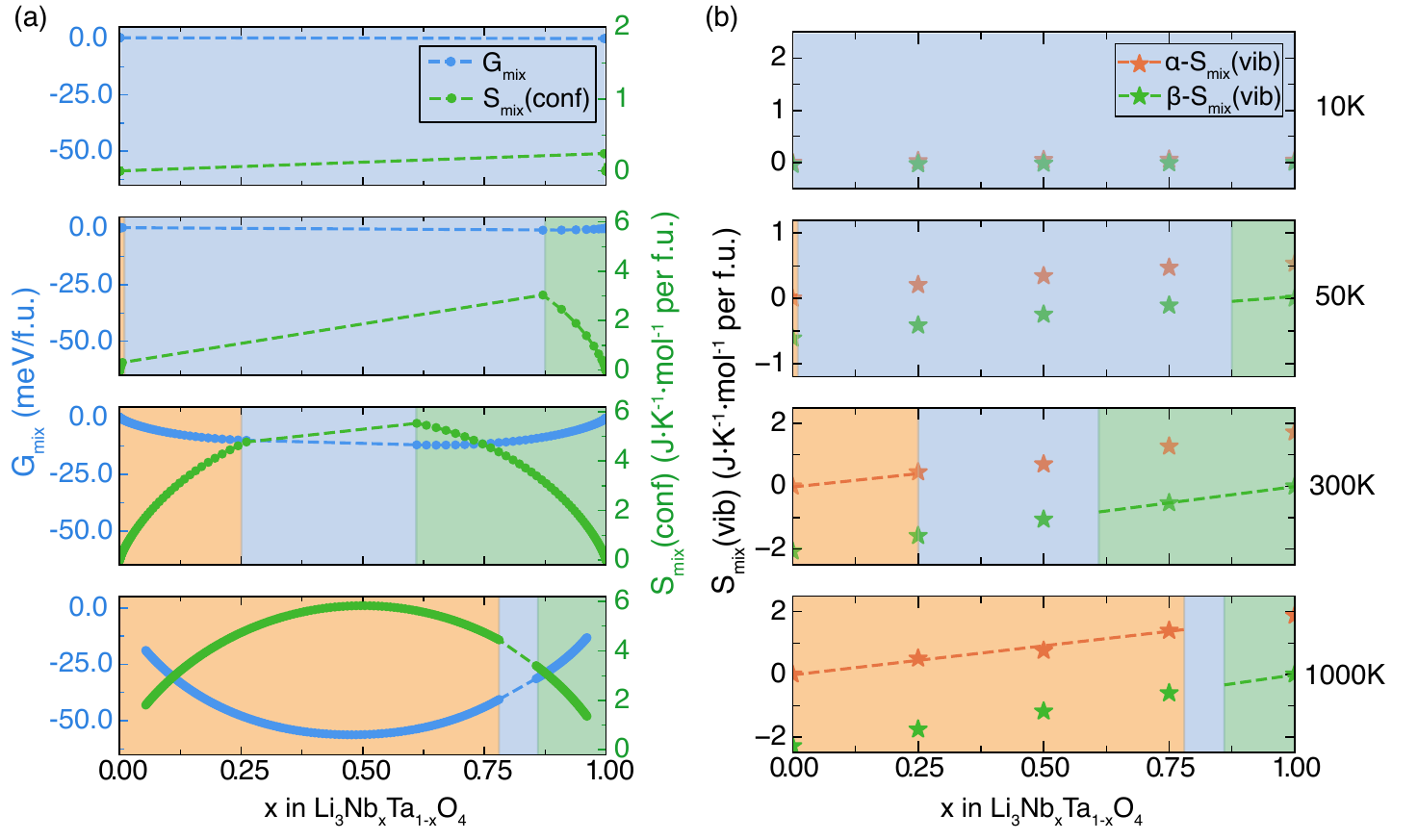}
    \caption{\label{Entropy}
    Thermodynamic properties (vertical axis) are shown as a function of Nb composition (x, horizontal axis). (a) Gibbs free energy $\mathrm{G_{mix}}$ (blue), configurational entropy of mixing $\mathrm{S_{mix}(conf)}$ (green), and (b) Vibrational entropy of mixing $\mathrm{S_{mix}(vib)}$ is plotted against composition ($x$) for the $\alpha$ (orange) and $\beta$ (green) phases at 10 K, 50 K, 300 K, and 1000 K. Stars denote the $\mathrm{S_{mix}(vib)}$ values calculated for the ground-state structures at these temperatures. The colored regions indicate the phase distributions of $\alpha$-phase (orange), $\alpha + \beta$ biphasic region (blue), $\beta$-phase (green) at corresponding temperatures.
    }
\end{figure*}

Figure \ref{Entropy} emphasizes the thermodynamic properties of \ce{Li_3Nb_xTa_{1-x}O_4}, where the colors represent the phase regions at different temperatures ($\alpha$ phase in orange, biphasic $\alpha+\beta$ region in blue, and $\beta$ phase in green). The left panels describe the total Gibbs free energy of mixing, $\mathrm{G_{mix}}$ (blue points), and the configurational entropy of mixing, $\mathrm{S_{mix}(conf)}$ (green points), computed at selected temperatures, 10 K, 50 K, 300 K, and 1000 K. The right panels illustrate the vibrational entropy of mixing $\mathrm{S_{mix}(vib)}$ as a function of composition ($x$) for the $\alpha$-phase (orange) and $\beta$-phase (green). Star symbols denote the calculated $\mathrm{S_{mix}(vib)}$ for the ground-state structures within each phase. 

As shown in Figure \ref{Entropy}a, the Gibbs free energy of mixing, $\mathrm{G_{mix} =E_{mix}- TS_{mix}}$, decreases with increasing temperature due to configurational and vibrational entropy contributions. As $x$ (i.e., the Nb concentration) increases from 0 to 1, $\mathrm{S_{mix}(conf)}$ initially increases in $\alpha$-phase region, reaching a maximum at $x$=0.5 due to complete Nb/Ta disorder (for example, at 1000 K), and then decreases in the $\beta$-phase region. The magnitudes of $\mathrm{S_{mix}(conf)}$ values are comparable though in different phases. In Figure \ref{Entropy}b, the vibrational entropies of mixing, $\mathrm{S_{mix}(vib)}$, show a consistent increase with increasing Nb concentration in  $\alpha$- and $\beta$-phases. Notably, negative values of $\mathrm{S_{mix}(vib)}$ are observed in the $\beta$-phase. Typically, in regimes of solid solution, negative values of vibrational entropy of mixing enhance thermal stability in ordered states, suppressing the formation of disordered structures at higher temperatures, leading to retrograde solubility.\cite{Benz1970retrogrades,Delaire2004Svib} This observation helps to explain the observed decrease in Ta solubility in the $\beta$-phase as temperature increases. Accounting for vibrational entropy differences (Figure \ref{pd_L3NTO4}b) between phases significantly corrects transition temperatures relative to estimates based solely on configurational entropy (Figure \ref{pd_L3NTO4}a).

\subsection{Li$^+$ Mobilities in \ce{Li_3Nb_xTa_{1-x}O_4}}

Prior experimental studies on partially-crystallized LiNb$_x$Ta$_{1-x}$O$_3$ reported enhanced Li-ion conductivities, exhibiting enhancements within one order of magnitude relative to the pure \ce{LiNbO_3} and \ce{LiTaO_3}, contingent on composition and synthesis protocols.\cite{Yada2015,LiuAl2023,Wang2020Unveiling} Thus, we explore if Li-ion conductivity enhancements also manifest in the mixed \ce{Li_3Nb_xTa_{1-x}O_4} system. 

Point defects, including vacancies, interstitials, and anti-site defects, as well as their concentrations, control Li-ion mobility in materials. Previous experimental characterizations and  computational reports \cite{Iyi1992,Xin2021LNL3Ncoating,Kim2023L3Tdefect,Chen2023NTcoating} suggested that \ce{LiMO_3} (M~=~Nb, Ta) incorporates defects deviating from its stoichiometry. These point defects are characterized by $\sim 1\%$ M antisites and $\sim 4\%$ Li vacancies. \citeauthor{Vyalikh2018} showed that Ta interstitials and Li vacancies are more likely to form in \ce{LiTaO_3}.\cite{Vyalikh2018} Furthermore, the coexistence of \ce{LiMO_3} and \ce{Li_3MO_4} phases in Nb(Ta)-lithium oxides is expected to favor defect formation within \ce{Li_3MO_4}, leading to comparable defect concentrations between the two phases. 

Stoichiometric \ce{Li_3TaO_4} and \ce{Li_3NbO_4} are poor ionic conductors due to their fully occupied Li sites.\cite{ruprecht2010ultraslow} While introducing defects, such as Li vacancies, is a known strategy to enhance conductivity,\cite{Kim2023L3Tdefect} the fundamental transport properties of these charge carriers are not well established.

To quantify the intrinsic ion mobility in these systems, here we computationally investigated the low-vacancy limit regime by introducing one Li-vacancy ($\sim1\%$) per formula unit. Charge neutrality of the simulation was achieved by introducing a compensating background charge. Based on the phase diagram of Figure~\ref{pd_L3NTO4}b, we selected configurations at various Nb concentrations ($\alpha$-phase: $x$~=~0, 0.125, and 0.1875; and $\beta$-phases: 0.6875, 0.75, and 1) from MC simulations at 300~K. The ionic conductivity of each structure was subsequently estimated by molecular dynamics using machine learned moment-tensor potentials (MTP-MD, see Section~{\ref{MTP-MD}}). The resulting conductivity, while modest, represents the intrinsic mobility in a dilute-defect scenario. However, it should be noted that the overall conductivity of a synthesized material is ultimately dictated by the total vacancy concentration generated by specific processing protocols. In practice, synthesis protocols are expected to generate a higher concentration of defects, leading to greater ionic conductivities than those predicted here.

\begin{figure}[ht!]
    \centering
    \includegraphics[width=1.0\columnwidth]{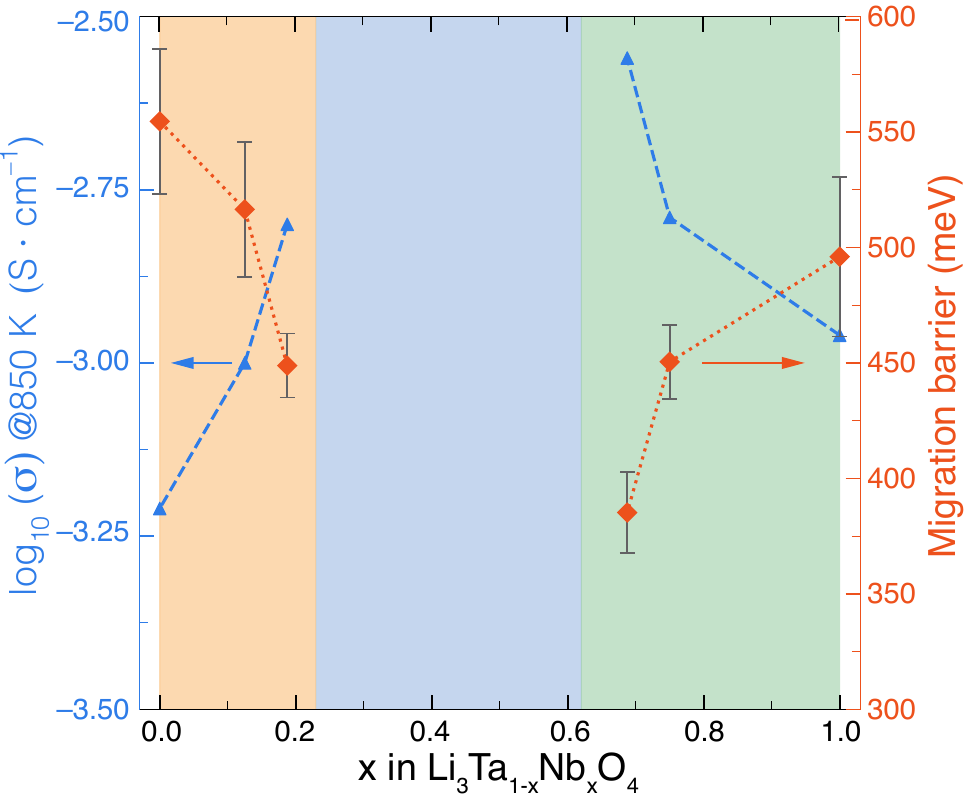}
    \caption{\label{MTP_Ea}
   Temperature-dependent ionic conductivities $\sigma$ (in S$\cdot$cm$^{-1}$) (at 850 K) and Arrhenius-fitted migration barriers (in meV) of \ce{Li_3Nb_{x}Ta_{1-x}O_{4}} for $x$~=~0, 0.125, 0.1875, 0.6875, 0.75, and 1 derived from machine learning MTP molecular dynamics calculations. Colored regions represent the phase distributions at 300~K: $\alpha$ phase, $x$=0.00--0.23, $\alpha$+$\beta$ phases, $x$=0.23--0.62, $\beta$ phase, $x$=0.62--1.00. The error bars of migration barriers fitted from the temperature-dependent Li$^+$ diffusivities are  shown. } 
\end{figure}

The calculated Li-ion transport properties for the $\ce{Li_3Nb_xTa_{1-x}O_4}$ system reveal that cation mixing enhances ionic conductivity by lowering the activation energy for Li-ion migration. As shown in Figure~\ref{MTP_Ea}, the migration barriers ($E_a$) decrease within both the Ta-rich $\alpha$-phase (orange area) and the Nb-rich $\beta$-phase (green area) as they approach the central, phase-separated area.

Note, we did not calculate the ionic conductivities of structures in regimes of phase separation ($x$=0.23-0.62, identified by the blue area in Figure~\ref{MTP_Ea}). However, recent kinetic Monte Carlo simulations of solid electrolytes in either solid solution or phase separation regimes indicate that the overall ionic conductivities of many compositions are unaffected by the underlying phase behavior.\cite{Deng2022KMC}

Specifically, in the $\alpha$-phase, the average migration energy decreases from 555$\pm$31 meV in pure \ce{Li_3TaO_4} ($x$=0) to 449$\pm$20 meV at an Nb concentration of $x$=0.125. Conversely, in the $\beta$-phase, the barrier decreases from 496$\pm$34 meV in pure \ce{Li_3NbO_4} ($x$=1) to 385$\pm$15 meV at $x$=0.6875. This reduction in activation energy directly translates to higher Li-ion conductivities at elevated temperatures. At 850 K, the computed conductivity of the $\alpha$-phase increases from 6.12$\times 10^{-4}$ to 1.60$\times 10^{-3}$~\ce{S\cdot cm^{-1}} as $x$ increases from 0 to 0.1875. Similarly, in the $\beta$-phase, the Li-ion conductivities increase as more Ta is introduced, rising from 1.10$\times 10^{-3}$ ($x$=1) to 2.78$\times 10^{-3}$ \ce{S\cdot cm^{-1}} ($x$=0.6875).

Our calculated migration energy barriers for the stoichiometric end-members show good agreement with existing experimental data. For $\alpha$-\ce{Li_3TaO_4}, our computed migration barriers of 555$\pm$31 meV (with ~1\% Li vacancies) match well with reported experimental values of $\sim$550~meV for Li-deficient samples (with $\sim$0.67\% Li vacancies) and 630-680 meV for stoichiometric samples.\cite{Kim2023L3Tdefect,Shari2020} For $\beta$-\ce{Li_3NbO_4} (with ~1\% Li vacancies), the calculated migration barriers of 496$\pm$34 meV are slightly lower than the experimentally observed values (580-850 meV) for crystalline and stoichiometric samples.\cite{RdeAzevedo2025,Glass1978Amorphous,ruprecht2010ultraslow}

Furthermore, the general trend of enhanced ionic transport upon mixing is consistent with findings in related material systems. For example, \citeauthor{Yada2015} investigated partially-crystallized Li-Nb-Ta oxides for battery interlayers and found that a mixed-cation composition (\ce{Li_{56}Nb_{22}Ta_{22}} oxide, $x$=0.5) exhibited a higher ionic conductivity (4.2 $\times 10^{-6}$~\ce{S\cdot cm^{-1}}) than end member compounds \ce{LiNbO_3} (1.8 $\times 10^{-6}$~\ce{S\cdot cm^{-1}}) it was compared against. This supports our observation that blending Nb and Ta can be a beneficial strategy for improving Li-ion kinetics in these coating materials.

\section{Discussion}
Niobium and tantalum oxides are widely used in applications demanding exceptional thermal stability and dielectric performance, including high-performance faradaic capacitors, high-temperature superalloys, non-linear optics, and protective coatings.\cite{Huband2017, Bartasyte2019, Bernhardt2024LNTODFT, Kofahl2024} Understanding the properties of functional oxide-based coating materials is central to the development of high-energy density, safe rechargeable lithium-ion batteries with improved cycle stability and shelf life. 

Although lithium-niobate-tantalate oxides have been implemented through trial-and-error procedures into commercial Li-ion batteries, there are just a handful of experimental investigations establishing a link between the improved battery figures of merit and the structure and transport properties of these oxides.\cite{Yada2015,Zhangwenbo2017,Ruesing2016vibrationalLNT} This work elucidates how mixing niobium and tantalum affects the phase behavior and connected properties of their respective lithium oxides.

Our previous investigation on Li-Nb-O and Li-Ta-O materials suggested that niobate and tantalate materials feature as a mixture of different phases, \ce{LiNbO_3} and \ce{Li_3NbO_4}, or \ce{LiTaO_3} and \ce{Li_3TaO_4}, in various proportions.\cite{Chen2023NTcoating} The weight fractions of these phases (once coated on the active particles of positive electrode materials) are entirely controlled by the ratio of precursors and the calcination temperatures adopted during their preparation.\cite{Xin2021LNL3Ncoating, YantaoZhang2014} 

Nonetheless, the correlation between the coexistence of Nb and Ta within a single coating matrix (\ce{LiNb_xTa_{1-x}O_3} and \ce{Li_3Nb_xTa_{1-x}O_4}) and its thermodynamic characteristics and Li-ion transport behavior remains insufficiently investigated. We bridge this gap using advanced cluster-expansion-based Monte Carlo simulations that explicitly account for configurational and vibrational entropy, enabling a rigorous thermodynamic description of these complex oxides. Starting from thermodynamically meaningful structures of mixed \ce{Li_3Nb_xTa_{1-x}O_4}, we have investigated the Li-ion transport characteristics in these coxides through state-of-the-art moment tensor machine learning potentials. 

Aside from the enhanced Li-ion transport properties achieved by greater Ta incorporation, mixing Ta into \ce{Li_3NbO_4} will contribute to a mechanically and chemically more stable structure given the shorter and stronger Ta-O bonds ($\sim$1.997 \AA, \ce{\Delta H_{formation}^{298K}=805~kJ/mol}) than Nb-O bonds ($\sim$2.038 \AA, \ce{\Delta H_{formation}^{298K}=753kJ/mol}).\cite{Dean1973Nb-Obond} The improved mechanical properties are demonstrated by \ce{LiTaO_3} possessing higher bulk and Young's moduli than \ce{LiNbO_3}.\cite{Wang2023L3NL3Tmechanical} \ce{LiTaO_3} has been shown to enhance the structural stability of positive electrode materials when coated on their particles.\cite{Gaillac2016elastic, HyoBinLee2022Surface} 

Computational investigations have shown that \ce{LiTaO_3} and \ce{Li_3TaO_4} possess a wider stability window (vs.\ Li/\ce{Li^{+}}) than mixture of \ce{LiNbO_3} and \ce{Li_3NbO_4}, respectively.\cite{Xiao2019coatingscreen,YizhouZhu2016Coatingscreen} A recent comparison of \ce{LiNbO_3} and \ce{LiTaO_3} coatings by \citeauthor{Lee2021comparison},\cite{Lee2021comparison} lithium tantalate coating materials appear superior in mitigating oxidative reactions between positive electrode materials and sulfide-based solid electrolytes. These distinctions prompt the additional potential advantages of Nb/Ta mixing coating materials for enhanced performance.\\

\subsection*{Structural and Vibrational Origins of Nb and Ta Mixing Energetics} 

\noindent Belonging to transition metals of the Group 5, Nb and Ta share similar electronic characteristics in their outer shells (Nb:\ $\ce{[Kr]}$ 4d$^4$5s$^1$, Ta:\ $\ce{[Xe]}$ 4f$^{14}$5d$^3$6s$^2$).  The similarities between niobium and tantalum extend to their pentavalent cations, Nb$^{5+}$ and Ta$^{5+}$, which, in six-fold coordination, possess identical ionic radii of  $\sim$0.64~\AA.\cite{Shannon1969,Shannon1976,Giacovazzo2012} Due to their identical charge and nearly identical ionic radii, Nb$^{5+}$ and Ta$^{5+}$  readily substitute for one another in a wide range of oxide and alloy systems—often referred to as bronzes—without inducing appreciable lattice distortion. We utilize both qualitative and quantitative information for Nb$^{5+}$ and Ta$^{5+}$ to elucidate our findings in \ce{Li_3Nb_xTa_{1-x}O_4}. 

In general, a close inspection of the DFT-computed Na/Ta orderings suggests shorter average Ta-O distances compared with Nb-O bond lengths (as shown in Figure S7, \ce{LiNb_xTa_{1-x}O_3}, and Figure S8, \ce{Li_3Nb_xTa_{1-x}O_4} of the SI), in agreement with previous studies.\cite{Ruesing2016vibrationalLNT,Boulay2003L3Tcif,Jacquet2017L3Ncif,Radhakrishnan2012bondstrength} At first glance, these observations stand in marked contrast to Pearson's Hard and Soft Acids and Bases theory,\cite{Pearson1963} as Ta$^{5+}$ is considered a slightly softer cation than Nb$^{5+}$ due to its greater polarizability and correspondingly lower charge density.  

Unlike \ce{Li_3NbO_4} ($I\overline{4}3m$), a high-symmetry cubic analog phase for \ce{Li_3TaO_4} has never been reported, indicating that octahedral coordination is less favorable for \ce{Ta^{5+}} within this structural framework.\cite{Blasse1965L3NTO4} \ce{Li_3NbO_4} features tetramers of Nb octahedron clusters, where each \ce{Nb^{5+}} has three nearest neighbours \ce{Nb^{5+}}s, with a high-density of pentavalent cations. In contrast, \ce{Ta^{5+}} ions are more dispersed in \ce{Li_3TaO_4}, which exhibits electrostatically stabilized zig-zag chains of Ta octahedra, where each \ce{Ta^{5+}} has only two  \ce{Ta^{5+}} nearest neighbors. 

Consequently, substituting Ta into \ce{Li_3NbO_4} may introduce substantial lattice strain, necessitating ``strong'' atomic vibrational coupling with the Ta site centers to preserve the structure. Although the specific vibrational modes of \ce{Li_3Nb_xTa_{1-x}O_4} alloys are not explicitly investigated in this work, we observed the decrease in vibrational entropy of mixing with increasing Ta content (decreasing Nb amount) in the $\beta$-phase (Figure~\ref{Entropy}). As a result, \ce{Li_3NbO_4} is less likely to mix large amounts of \ce{Ta^{5+}}, resulting in a progressively narrower region of $\beta$ phase as the temperature increases (Figure~\ref{pd_L3NTO4}b). On the contrary, the lower spatial density of \ce{Ta^{5+}} in \ce{Li_3TaO_4} leads to weaker cation--cation interactions, facilitating mixing of  \ce{Nb^{5+}} to stabilize the $\alpha$-phase.

This set of evidence reinforces the idea that including vibrational entropy effects in addition to configurational entropy contributions (as usually done through cluster expansion and Monte Carlo approaches) appears crucial to reproduce the miscibility of Ta in \ce{Li_3NbO_4}, as demonstrated in Figure~\ref{pd_L3NTO4}b. These findings underline the importance of incorporating vibrational entropy for accurate phase diagram predictions of complex materials,\cite{Neuhaus2014,Shao2023,Delaire2004Svib,Ozolins1998} especially when phases differ only subtly in configurational energy.

\subsection*{Li-ion mobility in \ce{\mathbf{Li_3Na_xTa_{1-x}O_4}}}

\noindent Previously, we have shown that crystalline \ce{Li_3NbO_4} (and \ce{Li_3TaO_4}) have higher Li-ion transport properties than crystalline \ce{LiNbO_3} (or \ce{LiTaO_3}),\cite{Chen2023NTcoating} with the \ce{Li_3MO_4} phase fraction forming preferentially with a rich defect landscape at higher calcination temperatures. Focusing on \ce{Li_3Na_xTa_{1-x}O_{4}} here, we have revealed that mixing Nb and Ta, i.e., substituting Ta with Nb in the $\alpha$-phase (or replacing Nb with Ta in the $\beta$-phase) is a possible way to enhance Li-ion conductivities of these compounds. 

The mechanism for the enhanced Li-ion conductivity upon mixing can be attributed to modifications of the energy landscape and local vibrational properties of the system. Introducing a guest cation creates localized strain and distorts the polyhedra that define the Li-ion diffusion pathways. In both $\alpha$ and $\beta$ phases, partial substitution creates local site-energy heterogeneity that reduces the correlation between adjacent Li-ion hops, lowering the average activation barrier and opening percolating low-barrier pathways. Such behavior reflects a general mechanism reported for disordered oxides, where cation mixing disrupts long-range order and broadens the distribution of accessible Li sites.

In addition to static disorder imposed by Ta and Nb mixing during synthesis, vibrational effects might also modulate transport properties differently across the two phases.  In the Ta-rich $\alpha$-phase, the negative mixing vibrational free energy (Figure S11) suggests a stabilization of lattice softening upon Nb substitution, consistent with a reduction in bottleneck size that would further lower migration barriers. Conversely, in the Nb-rich $\beta$-phase, substituting with Ta yields a positive mixing vibrational free energy (Figure S11). This reveals that while this mixed state is vibrationally destabilized relative to the pure components, it might paradoxically promote faster ion migration. A more comprehensive understanding of ionic transport properties necessitates future investigations that precisely quantify the interplay between configurational and vibrational contributions, thereby elucidating the roles of structural disorder and lattice dynamics in complex materials.

% We have attributed this increase in Li-ion transport on LNTO to the smaller volumes occupied by \ce{TaO_6} octahedra units as compared to more cumbersome \ce{NbO_6} analogs. However, at compositions $x \leq 0.6$, \ce{Li_3Ta_{1-x}Nb_{x}O_{4}}  dissolution of Ta into the $\beta$-phase is suppressed by a biphasic regime, and eventually into a conversion to the lower \ce{Li^+} conductivity $\alpha$-phase (achieved at higher Nb contents). Therefore, future studies should optimize the synthesis and calcination protocols to facilitate greater dissolution of Ta into the $\beta$-phase as a mean to increase Li-ion mobilities in LNTOs. 

\section{Conclusion}

In summary, we investigated the effect of niobium and tantalate mixing in lithium oxides, forming \ce{Li_3Nb_xTa_{1-x}O_4} alloys relevant to improving ionic conductivity and electrochemical performance in lithium-ion batteries. Using a multiscale approach parametrized on first-principles density functional theory calculations, we have successfully mapped the complex phase diagram of \ce{Li_3Nb_xTa_{1-x}O_4}, demonstrating qualitative agreement with reported phase transitions. 

Crucially, we unveiled the substantial influence of vibrational contributions on reducing the solubility of Ta in \ce{Li_3NbO_4}. Moreover, our study highlights the potential for enhancing Li-ion transport performance in crystalline \ce{Li_3Nb_xTa_{1-x}O_4} through strategic Nb and Ta mixing. Such mixing exhibits similar effects on different phases, leading to an improvement in Li-ion conductivity in the $\beta$-phase and $\alpha$-phase. These findings offer the potential to tailor ionic conductivities of \ce{Li_3Nb_xTa_{1-x}O_4} materials and optimize their performance as protective coatings in energy storage applications, high-performance faradaic capacitors, high-temperature superalloys, and non-linear optics.

 %%%%%%%%%%%%%%%%%%%%%%%%%%%%%%%%%%%%%%%%%%%%%%%%%%%%%%%%%%%%%%%%%%%%%
\begin{acknowledgement} 
P.C.\ acknowledges funding from the National Research Foundation under his NRF Fellowship NRFF12-2020-0012, Singapore.  The computational work was performed on the resources of the National Supercomputing Centre, Singapore (https://www.nscc.sg). P.C.\  acknowledges the Robert A. Welch Foundation under grants L-E-001-19921203, and additional financial support from the Welch Foundation under award E-2227-20250403.
. 
\end{acknowledgement} 
%%%%%%%%%%%%%%%%%%%%%%%%%%%%%%%%%%%%%%%%%%%%%%%%%%%%%%%%%%%%%%%%%%%%%

%%%%%%%%%%%%%%%%%%%%%%%%%%%%%%%%%%%%%%%%%%%%%%%%%%%%%%%%%%%%%%%%%%%%%
\begin{suppinfo}

The Supporting Information provides supplementary details including: Computational methodology and results of DFT calculations and Cluster Expansion fitting for both the \ce{Li_3Nb_xTa_{1-x}O_4} and \ce{LiNb_xTa_{1-x}O_3} systems; Tabulated M-O bond lengths and free volume ratios for all DFT-calculated ordered structures; Calculated phonon dispersion curves for the identified ground-state structures; A detailed description of the thermodynamic integration method employed in this work.

\end{suppinfo}
%%%%%%%%%%%%%%%%%%%%%%%%%%%%%%%%%%%%%%%%%%%%%%%%%%%%%%%%%%%%%%%%%%%%%

%%%%%%%%%%%%%%%%%%%%%%%%%%%%%%%%%%%%%%%%%%%%%%%%%%%%%%%%%%%%%%%%%%%%%%%%%%%%

\bibliography{references.bib}

%%%%%%%%%%%%%%%%%%%%%%%%%%%%%%%%%%%%%%%%%%%%%%%%%%%%%%%%%%%%%%%%%%%%%%%%%%%%
\end{document}